\documentclass[reprint,prb,twocolumn,showpacs,superscriptaddress,aps]{revtex4-1}

\usepackage{graphicx}
\DeclareGraphicsExtensions{.png,.jpg,.eps}

\usepackage{hyperref}
\usepackage{xcolor}
\usepackage{amsmath}

\begin{document}

\title{Quantum spin Hall effect in rutile-based oxide multilayers}

\author{J. L. Lado}
\affiliation{QuantaLab, International Iberian Nanotechnology Laboratory, Braga, Portugal}
\email{jose.luis.lado@gmail.com}

\author{Daniel Guterding}
\affiliation{Institut f\"ur Theoretische Physik, Goethe-Universit\"at Frankfurt,
Max-von-Laue-Stra{\ss}e 1, 60438 Frankfurt am Main, Germany}
\email{guterding@itp.uni-frankfurt.de}

\author{Paolo Barone}
\affiliation{Consiglio Nazionale delle Ricerche (CNR-SPIN), 67100 L'Aquila, Italy}
\affiliation{Graphene Labs, Istituto Italiano di Tecnologia, via Morego 30, 16163 Genova, Italy}

\author{Roser Valent\'i}
\affiliation{Institut f\"ur Theoretische Physik, Goethe-Universit\"at Frankfurt,
Max-von-Laue-Stra{\ss}e 1, 60438 Frankfurt am Main, Germany}

\author{V. Pardo}
\affiliation{Departamento de F\'{i}sica Aplicada,
Universidade de Santiago de Compostela, E-15782 Campus Sur s/n,
Santiago de Compostela, Spain}
\affiliation{Instituto de Investigaci\'{o}ns Tecnol\'{o}xicas,
Universidade de Santiago de Compostela, E-15782 Campus Sur s/n,
Santiago de Compostela, Spain}

\begin{abstract}
Dirac points in two-dimensional electronic structures are a source for topological electronic states due to the
$\pm \pi$ Berry phase that they sustain. 
Here we show that
two rutile multilayers (namely
(WO$_2$)$_2$/(ZrO$_2$)$_n$ and (PtO$_2$)$_2$/(ZrO$_2$)$_n$, where an active bilayer is sandwiched by a thick enough (n=6 is sufficient) band insulating substrate, show semi-metallic Dirac dispersions
with a total of four Dirac cones along the $\Gamma-M$ direction.
These become gapped upon the introduction of
spin-orbit coupling, giving rise
to an insulating ground state comprising four edge states. We discuss
the origin of the lack of topological protection
in terms of the valley spin-Chern numbers and the multiplicity of Dirac points.
We show with a model Hamiltonian that mirror-symmetry breaking
would be capable of creating a quantum phase transition
to a strong topological insulator, with a single Kramers pair per
edge.
\end{abstract}

\maketitle

\section{Introduction}

Topological states of matter\cite{RevModPhys.82.3045,RevModPhys.83.1057} 
have been the source of tremendous excitement and have fostered a
rich variety of new (and old) ideas in condensed
matter physics.
Quantum Hall effect, quantum spin Hall states, \cite{PhysRevLett.95.226801} topological crystals and topological crystalline insulators
are some examples in which the topology of the single-electron Hamiltonian translates
into robust electronic transport and surface states, resilient to the typical
perturbations that real samples will have, such as defects and impurities. The different types of quantum
Hall effects have provided us with a way to easily measure the quality
of samples and even the charge of the electron, whereas
quantum spin Hall states realize chiral wires,\cite{PhysRevLett.95.226801} 
where momentum
and spin are coupled together.
Furthermore, their superconducting
analogues, topological superconductors\cite{PhysRevLett.100.096407}
are known to give rise
to Majorana bound states.\cite{RevModPhys.87.137}  
Such many-body states can show
non-abelian braiding properties,
making them suitable to become building blocks for topological
quantum computing. Moreover, the nature of some of these
topological states in superconductors still needs to be 
understood.\cite{wang2016topological}
A common way to design topological superconductors
is precisely based on a quantum spin Hall state
proximized 
to a conventional superconductor,\cite{PhysRevLett.100.096407} 
turning quantum spin Hall states into
a key ingredient not only in spintronics,\cite{RevModPhys.87.1213} 
but also in topological
quantum computing.

Two main mechanisms need to be understood in order to design quantum spin Hall states.
The first one corresponds to band inversion, which relies on spin-orbit coupling (SOC)
altering the order of the $s-p$ orbital characters
in a band structure, whose best known
realization are HgTe/CdTe quantum wells.\cite{bernevig2006quantum} 
The modification can occur at the $\Gamma$ point,
so that SOC basically changes the parity of the highest occupied
band. The second mechanism is the opening of a protected Dirac point
by SOC.\cite{PhysRevLett.95.226801} 
Dirac points host protected $\pm \pi$ Berry phases that become
$\pm 1/2$ Chern numbers upon a gap opening, so that in the presence
of time-reversal symmetry a total spin-Chern number gives rise to
the quantum spin Hall state, whereas
with broken time reversal symmetry a quantum anomalous Hall state is realized.

The theoretical quest for new topological insulators
has involved 
semiconductors\cite{konig2007quantum,zhang2009topological},
transition metal oxides,\cite{PhysRevB.82.085111}
metal organic frameworks\cite{wang2013prediction,wei2016spin}
or optical lattices.\cite{li2013topological} 
Among them,
oxides offer a rich set of possibilities due to the
variety of materials that can be synthesized in a chemically stable form, together with the ease of fabrication in low-dimensional forms such as thin films or multilayers.
Predictions of both, quantum spin Hall and quantum anomalous Hall
effect have appeared in various oxides based
on perovskites,\cite{xiao2011interface,PhysRevB.88.155119,liang2013electrically} 
pyrochlores,\cite{PhysRevB.82.085111,PhysRevLett.103.206805} 
rutiles,\cite{PhysRevB.92.161115} corundum \cite{PhysRevB.92.235102}
or doped Kagome structure.\cite{guterding2016prospect}
%or even in bulk materials.\cite{PhysRevLett.108.106401}
 Many of the
proposals rely on an underlying hexagonal lattice, where Dirac points
are prone to appear at the $K,K'$ corners of the Brillouin zone. Rutiles, however,
 develop (semi) Dirac points, but due to their
tetragonal unit cell, the Dirac points show up at a certain
k-point in the $\Gamma-X$ direction.\cite{PhysRevLett.102.166803} 
Opening a gap at these points
with a time-reversal symmetry breaking has been shown to give rise
to a quantum anomalous Hall state\cite{PhysRevB.92.161115}
with a very small gap of about $1$ meV. Whether a quantum spin
Hall state in a rutile-based structure can be engineered or not in such a way is still an open
issue.

In this manuscript we address whether
a rutile material with time-reversal symmetry is able to develop
a quantum spin Hall state by opening a gap at the Dirac points
of its band structure by the SOC effect.
For that sake, we have designed two different rutile multilayers that show Dirac points
in their band structure, i.e.
(WO$_2$)$_2$/(ZrO$_2$)$_4$ 
and
(PtO$_2$)$_2$/(ZrO$_2$)$_4$.
 Both heterostructures become gapped
upon introduction
of SOC.
We show by means of their topological
invariants and the calculation of their edge states,
that both systems realize a 2D crystalline topological insulating state,
characterized by four in-gap surface states.
We discuss the origin of this insulating state
in terms of the spin valley Chern numbers, whose multiplicity
is determined by the symmetry of the unit cell.
We propose a model Hamiltonian to describe the system
 and we show that breaking an in-plane mirror symmetry can drive a transition towards a strong topological
insulator by compensating two of the Dirac points and leaving just two of them
with uncompensated spin valley Chern numbers.

\section{Gapped Dirac points in non-magnetic rutiles}

The multilayers proposed (see Fig. \ref{fig1}a) are based on the rutile
structure and grow along the (001) direction
 with the $a,b$ lattice parameters fixed to
those of the band insulating substrate and the c-axis and internal atomic
positions fully relaxed. We use as insulating substrate rutile ZrO$_2$ whose
bulk lattice parameters were calculated {\it ab initio} (see Methods section)
 yielding a= 4.93 \AA, which
was the used value. The only requisite that the substrate should have is to
provide a substantial gap where the $d$ electrons of the  active bilayer are
allowed to form the Fermi surface without mixing with the states of the
substrate. We have tried other substrates, e.g. TiO$_2$ would be a good
candidate in terms of the size of its band gap. However, the
band bending it
introduces will be so large that it would destroy the Dirac points.
TiO$_2$
with a slightly enlarged $a$ parameter would be fine for our purposes,
or even a very thin layer of ZrO$_2$ on top of TiO$_2$ would also do.
WO$_2$ on top
of ZrO$_2$ would be almost unstrained (a= 4.86 \AA), but PtO$_2$ would undergo
substantial strain since our calculations yields a= 4.59 \AA.
We have also tried other insulating substrates, such as SnO$_2$ and PbO$_2$, 
whose lattice
matching will be better than ZrO$_2$, but their very narrow gap destroys the
Dirac points by
mixing substantially with the active $d$ electrons of the bilayer.

%The
%electronically active
%layers are  two unit cells of (001) PtO$_2$ and
%WO$_2$, forming a tetra-layer system with electron count 5d$^6$ and 5d$^2	$, respectively.
%

\subsection{Bulk electronic structure}

\begin{figure}[t!]
 \centering
                \includegraphics[width=.5\textwidth]{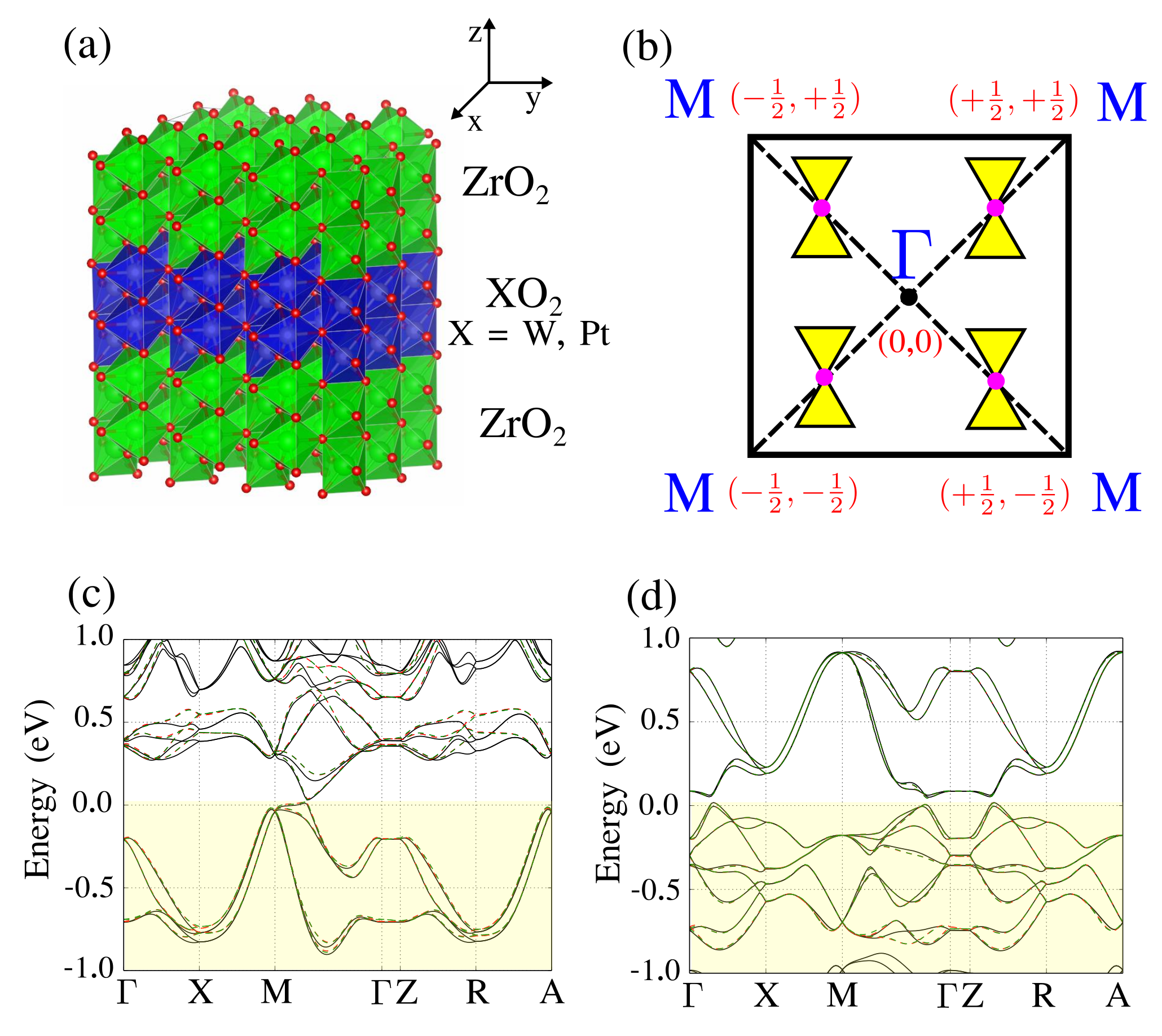}

\caption{ (a) Sketch of the multilayer structure, based on the
rutile unit cell, consisting of a bilayer of WO$_2$ or PtO$_2$
sandwiched between insulating ZrO$_2$. In the absence of
spin-orbit coupling, the low-energy  electronic properties
are dominated by four Dirac equations located along the
$\Gamma-M$ path, as shown in the sketch (b).
When SOC is introduced,
the band structure develops a gap for both the W-based (c) and Pt-based (d) multilayers.
The solid lines in (c,d) correspond to the DFT band structure,
whereas the dashed lines correspond to the Wannier interpolation.
In (d), the anti-crossing between $\Gamma$ and X is already present without SOC.
}
\label{fig:sketch}
\label{fig1}
\end{figure}

In the absence of SOC,
 a single crossing takes place along the $\Gamma-M$ direction
in both systems. Inspection
of the band dispersion around those points
reveals that the low-energy states form a Dirac cone.
When SOC is introduced, a gap opens up at the crossing points
(Fig. \ref{fig1}c,d), giving
rise to a bulk insulating state.

The topological properties of the band structure can be characterized
by calculating the topological invariant $\nu$ associated with
a system with time-reversal symmetry, by means of the $\ {Z}_2$
invariant. In the case
of crystals with inversion symmetry, it is known that the 
topological invariant can be obtained simply by calculating the parities of the
wavefunctions at the time
reversal invariant momenta (TRIM).\cite{PhysRevB.76.045302}
Nevertheless, the rutile structure presented here does not posses
global inversion symmetry, but only in-plane, so that we cannot apply
such procedure to the whole band structure. 
To calculate topological invariants for
systems without inversion symmetry several methods have been proposed:
(i) direct numerical
calculation of the Pfaffian,\cite{wimmer2012algorithm}
%explicit calculation of the Berry connection,
(ii) looking for an obstruction for a time-reversal smooth
gauge\cite{fukui2007quantum} and (iii) Wannier charge
center evolution.\cite{PhysRevB.83.235401,gresch2016z2pack} 
We have chosen this latter method, since it is numerically highly stable
and specially well suited for
electronic structure calculations. It is based on following the
flow of the Wannier charge centers as one moves across half of the Brillouin
zone. With this scheme, an odd number of crossings of the Wannier charge centers
label a non-trivial system, whereas an even number implies a trivial one.

From our calculations, we obtain that the
charge centers cross an even number of times, implying that
there is an even number of Kramers pairs per edge.
Nevertheless, we observe that the charge centers can be separated in
two families, each one showing an odd number of crossings, giving rise to 
a topological crystalline insulator.\cite{PhysRevLett.106.106802,gresch2016z2pack}
Every family of Wannier centers, whose existence is related
to the mirror symmetry of the unit cell, will produce a pair of Kramers
edge states, adding up to a total of four edge states.
The
gapless nature of these edge states
is not totally protected against perturbations even when
they respect time-reversal symmetry,
because
perturbations mixing the two families, such as chemical edge reconstruction,
will be able to open up an edge gap
This topological crystalline
insulating phase
is analogous to two coupled non-trivial quantum spin Hall states, as
in bilayer graphene with
spin-orbit coupling.\cite{prada2011band,PhysRevB.91.235451} 
Importantly,
topological
insulating states,
even if they are not protected against
symmetry mixing perturbations,
have been shown to be robust\cite{PhysRevLett.99.236809,sui2015gate,qiao2011electronic,li2016gate,li2011topological,PhysRevB.86.045102} 
against edge perturbations.

\subsection{Edge states}

\begin{figure}[t!]
 \centering
                \includegraphics[width=.5\textwidth]{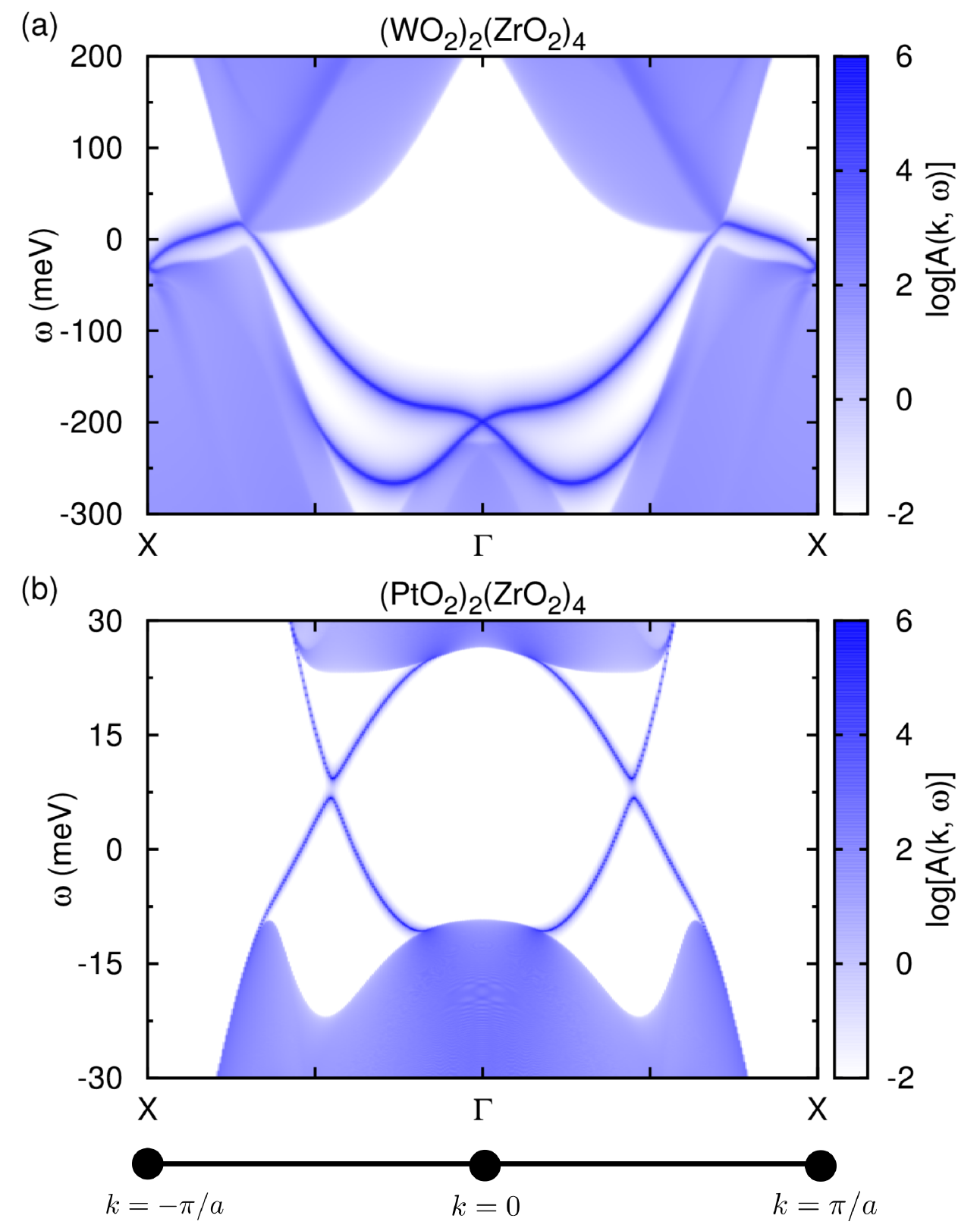}

\caption{
Edge (01) k-resolved density of states for the W-based (a) and Pt-based
multilayer (b). A set of edge states appear within the gap,
but due to the lack of a strong
topological index, their existence is not protected,
developing a small gap.
}
\label{fig:kdos}
\end{figure}

Having a bulk gapped spectrum and a crystalline topological invariant, the
rutile structures are expected to show edge states when studied in a finite
geometry. Using the Wannier Hamiltonian derived from the electronic
structure calculations, we calculate the surface spectral function
by solving Dyson's equation 

\begin{equation}
G(k_x,E) = (E - H_0 - t^\dagger(k_x)G(k_x,E)t(k_x))^{-1}
\end{equation}

where $H_0$ is the matrix with the intra-cell matrix elements, $k_x$
the vector parallel to the interface and $t(k_x)$ the hopping
of the Bloch Hamiltonian in the direction perpendicular to the
interface. From the Green's function, the surface spectral function is
calculated as $\Gamma(k_x,E) = \frac{1}{\pi} \text{Im} (G(k_x,E))$,
and shown in Fig. \ref{fig:kdos} for the W- and Pt- based rutile
multilayers.

The surface spectral function shows that both systems develop surface
states, a total of four of them.
%(as we sketched in Fig. \ref{fig1} b). 
The existence of four
edge states is coherent with the interpretation that the electronic structure
is equivalent to two coupled quantum spin Hall insulators.
In a topological quantum spin Hall insulator,
the number of surface states is 2 (or 6,8..), so that the surface hosts
a single Kramers pair, whose crossing cannot be avoided
due to time-reversal symmetry. In our present case, due to the
 existence of two Kramers pairs, perturbations can gap out the
surface states without breaking time-reversal symmetry. Kramers
degeneracy holds at the TRIM ($\Gamma$ and $X$), where the two branches
of edge states are degenerate in both valence and conduction band.

\section{Topological invariant in rutiles}

While the opening of Dirac points is a well known route to engineer
quantum spin Hall insulators, we have shown that in the present
rutile multilayers
quantum spin Hall states are realized that are not fully protected.
In terms of Wannier charge
centers, the origin
can be traced back to the additional symmetries of the system, namely mirror
symmetry,
that impose that there are two families of Wannier centers, each one
yielding a pair of Kramers edge states.
In terms of effective low energy Dirac points, the $C_4$ symmetry
imposes that the system has four spinless low energy Dirac points, yielding
twice the number of conventional honeycomb lattices, and twice as
many edge states.
Therefore, the combination of time reversal, $C_4$ and mirror symmetry,
and low energy effective Dirac Hamiltonian,
imposes that the system will show four edge states.
 In
the following we will try to give an intuitive understanding of why this
happens, as well as suggest a situation where a strong topological state
can be obtained by modifying the symmetry of our material.

\subsection{Origin of the topological insulating state}\label{origin}

Starting with the situation without spin-orbit coupling, the band structure
is characterized by the four Dirac crossings introduced at the beginning.
Independently of whether those crosses are Dirac, anisotropic Dirac
or semi-Dirac, the important feature for the present discussion is that
they carry a $\pm \pi$ Berry phase. We have verified numerically
that for the non-SOC calculations, the Wannier Hamiltonians obtained generate
 Dirac-like crosses with $\pm \pi$ Berry phase. The Dirac points
are located along the $\Gamma-M$ path (shown in Fig. \ref{fig1}b), where the position
depends on the details of the
electronic structure, giving rise to a total
of four non-equivalent Dirac points in the full Brillouin zone.

In the presence of SOC, a spin-dependent mass term appears
in the Hamiltonian. Due to the absence of inversion symmetry,
the eigenvalues are not degenerate except at
the TRIM points, where Kramer’s degeneracy is retained.
This is reflected
in the different masses
for different Kramers states. It has been shown that
the $\ {Z}_2$ invariant can be related to the so-called
spin Chern number, or simply to the Chern number $C$ of one of the Kramers
sectors. This can be easily understood in topological
states where $s_z$ is conserved, and the $ {Z}_2$ invariant
is 
\begin{equation}
\nu = C_{\uparrow} (\text{  mod } 2)
\end{equation}

which in the case of monolayer graphene gives $C_S = 2$,
$C_{\uparrow} = 1$
and $\nu=1$, while
for bilayer graphene it results in $C_S=4$, $C_{\uparrow}=2$ and
$\nu = 0$. In the case of rutile multilayers, due to the existence of
four gapped Dirac equations for each Kramers manifold, the Chern
number for a certain Kramers manifold will be a sum over the
Chern numbers of four Dirac equations $C_i$, which due to the
$\pm\pi$ Berry phase will be $C_i = \pm 1/2$. By labeling the
Kramers manifold as $\uparrow$ in analogy to the spin conserving case,
we have

\begin{equation}
C_{\uparrow} = \sum_{i=1}^4 s_i \frac{1}{2}
\end{equation}

where $s_i=\pm 1$ depending on the sign of the SOC-induced Dirac masses. 
Due to the in-plane inversion symmetry, the Chern number of the different Dirac
equations around $+\vec k$ and $- \vec k$ will be the same. 
We will have
that $C_1 = C_3 = s_1 \frac{1}{2}$ and $C_2=C_4= s_2 \frac{1}{2}$.
Therefore,
the Chern number for one of the Kramers sectors will be

\begin{equation}
C_{\uparrow} = s_1 + s_2 = 0, \pm 2
\end{equation}

With the previous Kramers Chern number, the topological invariant
will result into $\nu = 0$, so that the system will not be a
strong topological insulator in any case. For the case of
$C_{\uparrow} = +2$, time reversal symmetry ($C=0$) imposes that the
Chern number of the other sector is $C_{\downarrow} = -2$,
giving a total $C_S = 4$, predicting four edge states. This last
situation is precisely the one we obtained when calculating the
edge spectra of the Pt- and W- multilayers. 

We would like to emphasize that in the case
of having a similar system but with magnetic order, so that a single spin flavor is present at
the Fermi level, the previous argument will predict
that the total Chern number is 2.
This has been explicitly calculated in a V-based magnetic rutile 
multilayer\cite{PhysRevB.92.161115} that shows four semi-Dirac
points along $\Gamma-M$, obtaining that the
system is an anomalous Hall insulator showing two co-propagating
edge states.\cite{PhysRevB.92.161115}
It is interesting to note that, contrary to other
oxide systems,\cite{PhysRevB.89.195121} in this system magnetism brings about
the topological protection for the edge states, as
time-reversal symmetry breaking
leads to protected topological states.

The crystalline topological state in the
time reversal Pt- and W- multilayers is therefore a consequence of the
existence of a total of four Dirac points in the non-relativistic
band structure. This compares with the case of honeycomb lattices, where
usually only two Dirac equations are present, and the Chern
number spin flavor is $C_{\uparrow,\downarrow} = \pm 1$. 
In these rutile-based nanostructures, $C_4$ symmetry imposes that
a total of 4 Dirac equations appear between $\Gamma$ and $M$. If
such symmetry is broken, it could be possible to
have an electronic structure with only two Dirac equations that
could give rise to a strong topological insulating phase.
A way to realize that will be to expand
the cell along the (11) direction. With such distortion, two of the
Dirac points 
will displace differently than the other in k-space.
In particular, they can open up a trivial gap, whereas the
other two remain gapless (or with a much smaller gap).
In that situation, switching on SOC might be able to create
band inversion in two of the Dirac points, but not in the
other two, giving rise to $C_{\uparrow} = 1$ and robust quantum
spin Hall state $\nu = 1$.

\begin{figure}[t!]
 \centering
                \includegraphics[width=.5\textwidth]{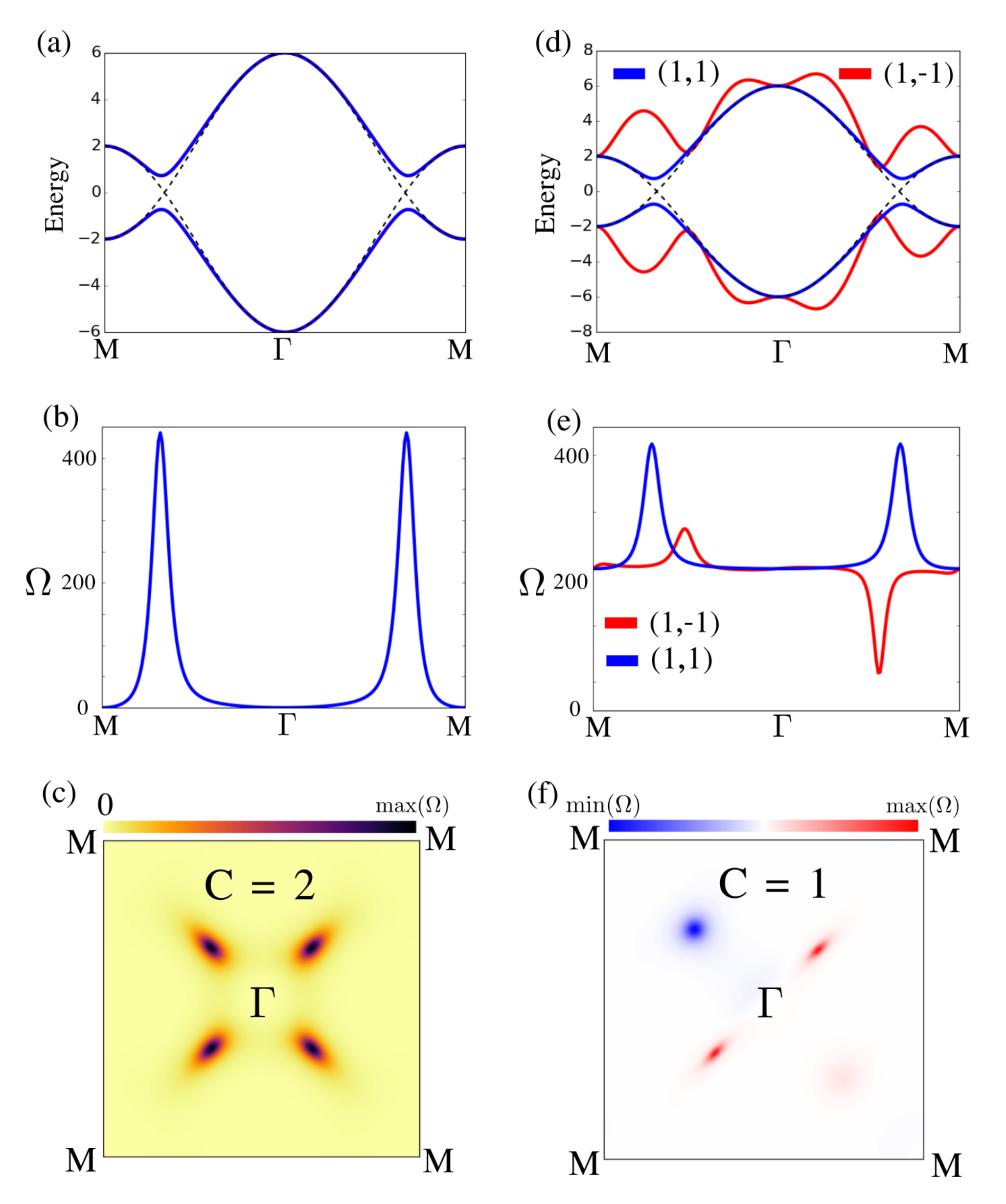}

\caption{ 
Band structure (a)
along the (1,1) direction for the model in Eq. \ref{eq:toy}, in the
quantum anomalous ($C=2$) regime ($\alpha=0$,$\lambda\ne0$), Berry curvature
along the path (b) and in the whole Brillouin zone (c). Dashed lines in (a)
show the bands for ($\alpha=\lambda=0$), the gapless regime.
Band structure (d), and Berry curvatures (e,f) in the
QAH regime with $\alpha\ne0,\lambda\ne0$, with total Chern number
$C=1$. Parameters are $\epsilon = -2$, $t=1$, $t'=2$,$\lambda=0.5$ and $\alpha=2$
}
\label{fig:toy}
\end{figure}

\subsection{Model Hamiltonian for spinless fermions}

In the following we will illustrate the
different topological states of the rutile lattice
by means of a model Hamiltonian.
We stress that the following model Hamiltonians do not intend
to precisely capture the electronic structure of the
two multilayers presented above, their goal is just
to provide an intuitive
understanding of the topology of the low-energy
electronic structure.
We choose a variation from a model previously shown to develop
anisotropic Dirac points\cite{PhysRevLett.103.016402} in a similar system.

\begin{equation}
H_{\uparrow} (\vec k) =
\begin{pmatrix}
\epsilon_1 (\vec k) & T (\vec k)\\
T^* (\vec k)& \epsilon_2(\vec k) \\
\end{pmatrix}
\label{eq:toy}
\end{equation}

with

\begin{gather*}
\epsilon_1 (k_x,k_y) = -\epsilon + 2t (\cos k_x +\cos k_y) \\
\epsilon_2 (k_x,k_y) = \epsilon - 2t (\cos k_x +\cos k_y) \\
T(\vec k) = V(\vec k) + W(\vec k) + Z (\vec k) \\
V(k_x,k_y) = 2t' (\cos k_x - \cos k_y ) \\
W(k_x,k_y) = 2 i\alpha \sin (k_x - k_y) \\
Z(k_x,k_y) = 2 i\lambda \sin k_x \sin k_y
\end{gather*}

The terms $\epsilon_1,\epsilon_2, V$ correspond to a simplified
version of anisotropic Dirac points.\cite{PhysRevLett.103.016402}
The term $W$ breaks the equivalence
between the $(1,1)$ and $(1,-1)$ directions but conserves time reversal.
The term $Z$ 
breaks time-reversal symmetry, but maintains the equivalence
between the (1,1) and (1,-1) directions.
In the case of $Z=W=0$, the previous Hamiltonian describes
a band structure that shows four anisotropic Dirac points, located
along the $\Gamma-M$ directions. The Dirac nodes
are equally spaced from $\Gamma$ and occur at points
$\vec k_0 = (\kappa_1,\kappa_2) k_0$, with $\kappa_{1,2} = \pm 1$

We first focus on the case where $W=0$ ($\alpha=0$),
but with a non-zero $Z$ ($\lambda \ne 0$) 
so that time-reversal symmetry is broken.
In this situation, the band structure develops a gap (Fig. \ref{fig:toy}a),
and the split Dirac points generate a non-zero Berry
curvature (Fig. \ref{fig:toy}b). When integrated over the whole
Brillouin zone, a Chern number $C=2$ is obtained. This situation
is analogous to the one observed\cite{PhysRevB.92.161115} 
in half-metallic V-based
multilayers with SOC, a system that is also a QAH with $C=2$,
but with the difference that the low-energy electronic structure is of type $II$ semi-Dirac\cite{PhysRevB.92.161115} 
instead of the anisotropic Dirac dispersion shown in Eq. \ref{eq:toy}.

If we first switch on the mirror-symmetry-breaking
term $W$, the spectrum consists of two gapped Dirac equations
in the (1,-1) direction
with opposite Chern numbers, and two gapless Dirac equations in the
(1,1) direction. If now time-reversal symmetry is broken by switching on $Z$
($\lambda\ne0$), and provided $\lambda$ is not large
enough to invert the gaps in (1,-1),
the Dirac equation in the (1,1) will open up a gap with the
same Chern number. The band structure
for this situation is shown in Fig. \ref{fig:toy}d, where
it is observed that the calculated Berry curvatures (Fig. \ref{fig:toy}e)
and Chern number (Fig. \ref{fig:toy}f) are in agreement
with the previous argument. Therefore, the model proposed in Eq. \ref{eq:toy}
shows a phase with Chern number $C=1$, provided mirror symmetry is broken.

The previous phenomenology can be understood by expanding the
Hamiltonian around the Dirac points.

\begin{equation}
H_\kappa = 
-\sigma_z p_1 + \kappa_1\kappa_2 \sigma_x p_2 -\sigma_y m_{\kappa}
\end{equation}

where $m_{\kappa}$ is a mass term induced by $W$ and $Z$,
$p_{1} = \kappa_1 p_x + \kappa_2 p_y$,
$p_{2} = -\kappa_2 p_x + \kappa_1 p_y$, with
$p_{x,y}$ the crystal momenta around the different valleys.
The Chern
number for the previous valley Hamiltonian
can be written as

\begin{equation}
\ {C}_{\kappa} = \frac{1}{2} \text{sign} ({\kappa_1\kappa_2 m_{\kappa}})
\end{equation}

In the case of $\lambda\ne0$ and $\alpha=0$, we have
$m_{\kappa} = \kappa_1\kappa_2 m$ so $\ {C}_{\kappa} = 1/2$,
and when summing over the four valleys it gives $\ {C} = 2$.
In the case of $\alpha\ne0$, two of the valley Chern numbers
can yield opposite signs, so that the net Chern number will be
$\ {C} = 1$.

\subsection{Model Hamiltonian for quantum spin Hall}

\begin{figure}[t!]
 \centering
                \includegraphics[width=.5\textwidth]{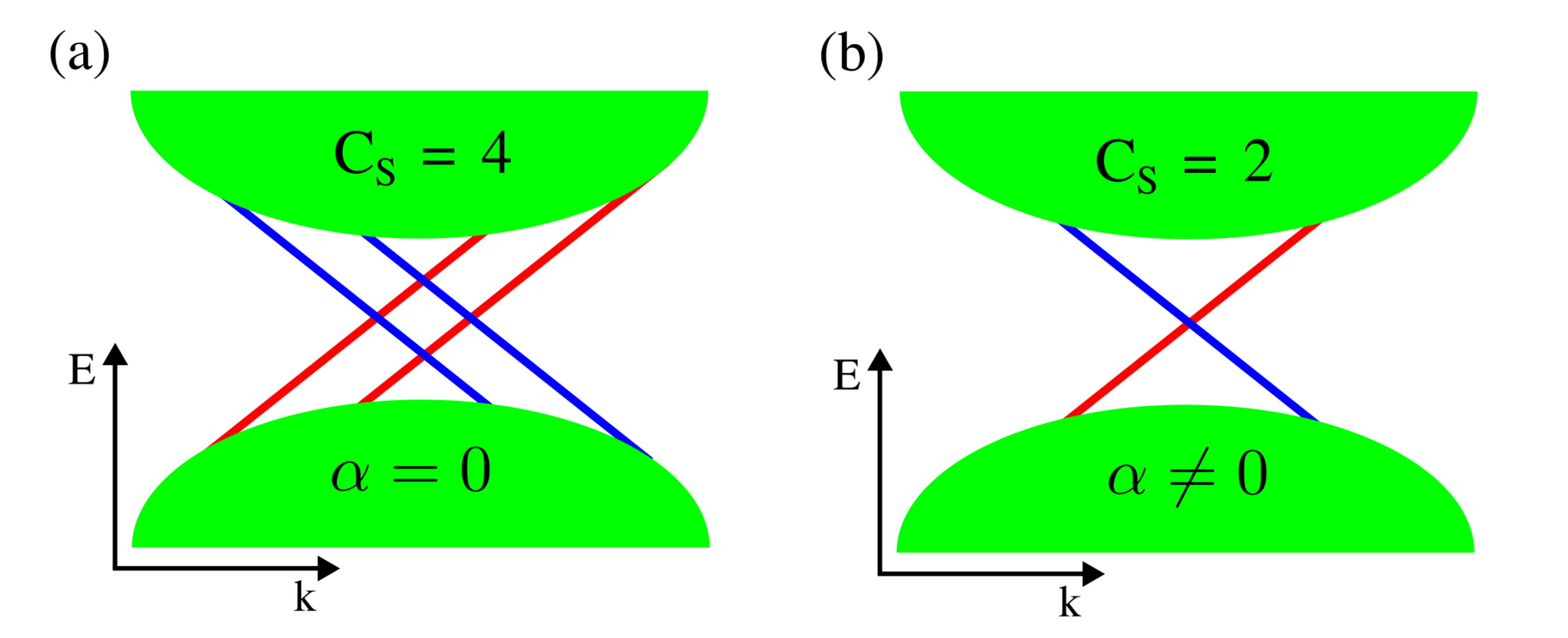}

\caption{ Sketch of the
edge states for the Hamiltonian Eq. \ref{ham_qsh}, showing the
crystalline insulating state with spin Chern number $C_S = 4$ (a), and the
strong phase with $C_S = 2$. The transition from (a) to (b)
is driven by the mirror symmetry parameter $\alpha$. In a real material,
such transition can be induced by strain in the (11) direction.
The DFT results of Fig. \ref{fig:kdos} correspond to the symmetric case (a).
}
\label{fig:qsh_toy}
\end{figure}

So far we have been studying a model for spinless fermions, showing
that it leads to two different quantum anomalous Hall states. With the
previous model
a spinful time-reversal version can be built as

\begin{equation}
H (\vec k)= 
\begin{pmatrix}
H_{\uparrow} (\vec k) & 0 \\
0 & H_{\downarrow} (\vec k) \\
\end{pmatrix}
=
\begin{pmatrix}
H_{\uparrow} (\vec k) & 0 \\
0 & H^*_{\uparrow} (- \vec k) \\
\end{pmatrix}
\label{ham_qsh}
\end{equation}

With the previous ansatz the net Chern number is zero as imposed by
time reversal, and the parameter $\lambda$ can be now understood as arising
from SOC. For $\alpha=0$, following the discussion in Sec. \ref{origin}, the
spin Chern number yields $C_S = 4$, giving rise to a crystalline topological
insulator
(Fig. \ref{fig:qsh_toy}a), compatible with the DFT results presented in Fig. \ref{fig1}.
Switching on the mirror symmetry term $\alpha$, the spin Chern
number yields $C_S = 2$, giving rise to a strong
topological insulator
(Fig. \ref{fig:qsh_toy}b).
Therefore, a mirror-symmetry breaking term
in our model Hamiltonian is capable of creating
a quantum phase transition from the original crystalline
topological insulating state
to a strong topological insulator.  
We finally clarify that in the real materials introduced in this
manuscript, spin mixing terms would show up in their effective Hamiltonian.

\section{Conclusions}
We have shown that two rutile-based bilayers formed by
an active 5d-electron system with 5d$^2$ and 5d$^6$ electron count
host a crystalline topological insulating state. The origin of the
non robust topological state comes from having four Dirac equations
in the absence of SOC, in comparison
with honeycomb lattices (whether this is graphene or oxide-based) that show only two.
This limitation can be traced to the Chern number per Kramers sector,
that in the rutile structure is 2 whereas in the honeycomb lattice is 1.
 We have
suggested that by removing undesired Dirac points, it would be possible
to design strong quantum spin Hall insulators in the rutile lattice.
We have shown with a toy model calculation
for spinless fermions 
that apart from
the insulating phase with 
$C=2$, breaking in-plane mirror symmetry will allow
to enter a state with $C=1$. The extension 
of the previous model with mirror symmetry breaking to
the spinful time-reversal case
would give rise to a strong 2D topological insulator.

\section*{Acknowledgments} 
J.L.L. acknowledges financial support from 
Marie-Curie-ITN
607904  SPINOGRAPH.
D.G and R.V. thank the Deutsche Forschungsgemeinschaft for funding
through SFB/TR 49.
V.P. thanks the
Xunta de Galicia for financial support under the
Emerxentes Program via Project No. EM2013/037 and the
MINECO via Project No. MAT2013-44673-R. V.P. acknowledges
support from the MINECO of Spain via the Ramon y
Cajal program under Grant No. RyC-2011-09024.
P.B. acknowledges partial support from the European Union’s Horizon 2020
research and innovation programme under Grant Agreement No. 696656
GrapheneCore.
We thank GEFES2016 for providing the platform for this collaboration to succeed.

\appendix

\section{Methods}

We have carried out density functional theory (DFT) calculations with various
codes: WIEN2k,\cite{schwarz2002electronic}  and Quantum Espresso\cite{giannozzi2009quantum} for
the various cell and geometry relaxations plus the electronic structure
analysis (both codes yielding comparable results), and using the code
FPLO\cite{fplo} for relativistic Wannierization calculations.
Structural relaxations using both WIEN2k and Quantum Espresso 
were carried out with the PBE version of the generalized gradient approximation\cite{PhysRevLett.77.3865} as an exchange-correlation functional, using PAW
pseudopotentials in the QE case and a full-potential calculation with WIEN2k,
both without SOC. The construction of Wannier functions was performed within the full-relativistic version of FPLO using a $6 \times 6 \times 3$ $k$-point grid. For the (WO$_2$)$_2$/(ZrO$_2$)$_n$ case we included all Zr $4d$ and W $5d$ states. For the (PtO$_2$)$_2$/(ZrO$_2$)$_n$ case we included all Zr $4d$, O $2p$ and Pt $5d$ states.

\bibliographystyle{ieeetr}
\bibliography{biblio}

\end{document}